\documentclass[3p,preprint,times,a4paper]{elsarticle}

\usepackage{amssymb}

\usepackage{graphicx}
\usepackage{dcolumn}
\usepackage{bm}
\usepackage{epsfig}
\usepackage{amsfonts, amsmath}
\usepackage{keyval,graphicx}
\usepackage{textcomp,wasysym}
\usepackage{subfigure}
\usepackage{xcolor}
\usepackage{longtable}
\usepackage{multirow}
\usepackage{epstopdf}
\usepackage{xspace}

\newcommand{\un}[1]{\ensuremath{\,\mathrm{#1}}}
\def \mev        {\ensuremath{\un{MeV}\xspace}}
\def \mevc       {\ensuremath{\un{MeV}/c}\xspace}

\def \gev        {\ensuremath{\un{GeV}\xspace}}
\def \gevc       {\ensuremath{\un{GeV}/c}\xspace}

\def \invpb      {\ensuremath{\un{pb^{-1}}}\xspace}

\def \nb         {\ensuremath{\un{nb}}\xspace}
\def \pb         {\ensuremath{\un{pb}}\xspace}

\journal{Phys. Lett. B}

\begin{document}

\begin{frontmatter}

\title{Study of $e^+e^- \rightarrow p\bar{p}$ in the
  vicinity of $\psi(3770)$ }

\author{%
\small
M.~Ablikim$^{1}$, M.~N.~Achasov$^{8,a}$, X.~C.~Ai$^{1}$, O.~Albayrak$^{4}$, M.~Albrecht$^{3}$, D.~J.~Ambrose$^{41}$, F.~F.~An$^{1}$, Q.~An$^{42}$, J.~Z.~Bai$^{1}$, R.~Baldini Ferroli$^{19A}$
% ...
\\
\vspace{0.2cm}
(BESIII Collaboration)\\
\vspace{0.2cm} {\it
$^{1}$ Institute of High Energy Physics, Beijing 100049, People's Republic of China\\
% ...
$^{3}$ Bochum Ruhr-University, D-44780 Bochum, Germany\\
$^{4}$ Carnegie Mellon University, Pittsburgh, Pennsylvania 15213, USA\\
% ...
$^{8}$ G.I. Budker Institute of Nuclear Physics SB RAS (BINP), Novosibirsk 630090, Russia\\
% ...
$^{19}$ (A)INFN Laboratori Nazionali di Frascati, I-00044, Frascati, Italy; (B)INFN and University of Perugia, I-06100, Perugia, Italy\\
% ...
$^{41}$ University of Rochester, Rochester, New York 14627, USA\\
$^{42}$ University of Science and Technology of China, Hefei 230026, People's Republic of China\\
% ...
$^{a}$ Also at the Novosibirsk State University, Novosibirsk, 630090, Russia\\
% ...
}
\vspace{0.4cm}
}

\begin{abstract}
Using $2917 \invpb$ of data accumulated at $3.773\gev$,
$44.5\invpb$ of data accumulated at $3.65\gev$ and data
accumulated during a $\psi(3770)$ line-shape scan with the BESIII
detector, the reaction $e^+e^-\rightarrow p\bar{p}$ is studied
considering a possible interference between resonant and continuum
amplitudes. The cross section of
$e^+e^-\rightarrow\psi(3770)\rightarrow p\bar{p}$,
$\sigma(e^+e^-\rightarrow\psi(3770)\rightarrow p\bar{p})$, is found
to have two solutions, determined to be $(0.059^{+0.070}_{-0.020}\pm0.012)\pb$
with the phase angle $\phi = (255.8^{+39.0}_{-26.6}\pm4.8)^\circ$ ($<0.166 \pb$
at the 90\% confidence level), or
$\sigma(e^+e^-\rightarrow\psi(3770)\rightarrow p\bar{p}) =
(2.57^{+0.12}_{-0.13}\pm0.12)\pb$ with $\phi = (266.9^{+6.1}_{-6.3}\pm0.9)^\circ$
both of which agree with a destructive interference. Using the
obtained cross section of $\psi(3770)\rightarrow p\bar{p}$, the
cross section of $p\bar{p}\rightarrow \psi(3770)$, which is useful
information for the future PANDA experiment, is estimated to be
either $(9.8^{+11.8}_{-3.9})\nb$ $(<27.5\nb$ at 90\% C.L.) or
$(425.6^{+42.9}_{-43.7})\nb$.

\end{abstract}


\end{frontmatter}


\section{Introduction}

At $e^+e^-$ colliders, charmonium states with $J^{PC}=1^{--}$, such
as the $J/\psi$, $\psi(3686)$, and $\psi(3770)$, are produced
through electron-positron annihilation into a virtual photon.
These charmonium states can then decay into light hadrons through
either the three-gluon process ($e^+e^-\rightarrow \psi \rightarrow
ggg \rightarrow hadrons$) or the one-photon process
($e^+e^-\rightarrow \psi \rightarrow \gamma^* \rightarrow hadrons$).
In addition to the above two processes, the non-resonant process
($e^+e^-\rightarrow \gamma^* \rightarrow hadrons$) plays an
important role, especially in the $\psi(3770)$ energy region where
the non-resonant production cross section is comparable to the
resonant one.

The $\psi(3770)$, the lowest lying $1^{--}$ charmonium state above
the $D\bar{D}$ threshold, is expected to decay dominantly into the
OZI-allowed  $D\bar{D}$ final states~\cite{MARK, DELCO}.
However, assuming no interference effects between resonant and non-resonant amplitudes, the BES Collaboration found a large total
non-$D\bar{D}$ branching fraction of
$(14.5\pm1.7\pm5.8)\%$~\cite{BES_nonDDbar_1,BES_nonDDbar_2,BES_nonDDbar_3,BES_nonDDbar_4}.
A later work by the CLEO Collaboration, which included interference
between one-photon resonant and one-photon non-resonant amplitudes
(assuming no interference with the three-gluon amplitude), found a
contradictory non-$D\bar{D}$ branching fraction of
$(-3.3\pm1.4^{+6.6}_{-4.8})\%$~\cite{CLEO_nonDDbar}. These different
results could be caused by interference effects. Moreover, it has
been noted that the interference of the non-resonant (continuum)
amplitude with the three-gluon resonant amplitude should not be
neglected~\cite{interfere_wangp_1}. To clarify the situation, many
exclusive non-$D\bar{D}$ decays of the $\psi(3770)$ have been
investigated~\cite{non_DDbar_exclucive_1, non_DDbar_exclucive_2}.
Low statistics, however, especially in the scan data sets have not
permitted the inclusion of interference effects in these exclusive
studies.

BESIII has collected the world's largest data sample of $e^+e^-$
collisions at $3.773\gev$. Analyzed together with data samples
taken during a $\psi(3770)$ line-shape scan, investigations of
exclusive decays, taking into account the interference of resonant
and non-resonant amplitudes are now possible. Recently, the decay
channel of $\psi(3770)\rightarrow
p\bar{p}\pi^0$~\cite{ppbarpi0_matthias} has been studied considering
the above mentioned interference. In this Letter, we report on a
study of the two-body final state $e^+e^- \rightarrow p\bar{p}$ in the
vicinity of the $\psi(3770)$ based on data sets collected with the
upgraded Beijing Spectrometer~(BESIII) located at the Beijing
Electron-Positron Collider~(BEPCII)~\cite{BESIII_BEPCII}. The data
sets include $2917\invpb$ of data at $3.773\gev$,
$44.5\invpb$ of data at $3.65\gev$~\cite{Lumi}, and data
taken during a $\psi(3770)$ line-shape scan in the energy range from
$3.74$ to $3.90\gev$.


\section{BESIII detector}

The BEPCII is a modern accelerator featuring a multi-bunch double
ring and high luminosity, operating with beam energies between 1.0
and $2.3\gev$ and a design luminosity of
$1\times10^{33}\un{cm^{-2}\,{s}^{-1}}$. The BESIII detector is a
high-performance general purpose detector. It is composed of a
helium-gas based drift chamber~(MDC) for charged-particle tracking
and particle identification by specific ionization $dE/dx$, a plastic scintillator
time-of-flight (TOF) system for additional particle identification,
a CsI~(Tl) electromagnetic calorimeter~(EMC) for electron
identification and photon detection, a super-conducting solenoid
magnet providing a 1.0~Tesla magnetic field, and a muon detector
composed of resistive-plate chambers. The momentum resolution for
charged particles at $1\gevc$ is $0.5\%$. The energy resolution
of $1\gev$ photons is $2.5\%$. More details on the accelerator
and detector can be found in Ref.~\cite{BESIII_BEPCII}.

A {\sc geant4}-based \cite{geant4}
Monte Carlo~(MC) simulation software package,
which includes a description of the geometry, material, and response
of the BESIII detector, is used for detector simulations. The signal
and background processes are generated with dedicated models that
have been packaged and customized for BESIII~\cite{generator}.
Initial-state radiation~(ISR) effects are not included at the
generator level for the efficiency determination, but are corrected
later using a standard ISR correction procedure~\cite{isr_1, isr_2}.
In the ISR correction, {\sc phokhara}~\cite{phokhara} is used to
produce a MC-simulated sample of $e^+e^-\rightarrow \gamma_{\rm
ISR}p\bar{p}$ (without $\gamma_{\rm ISR} J/\psi$ and $\gamma_{\rm
ISR} \psi(3686)$). For the estimation of backgrounds from
$\gamma_{\rm ISR} \psi(3686)$ and $e^+e^-\rightarrow
\psi(3770)\rightarrow D\bar{D}$, MC-simulated samples with a size
equivalent to 10 times the size of data samples are analyzed.

\section{Event selection}

The final state in this decay is characterized by one proton and one
antiproton. Two charged tracks with opposite charge are required.
Each track is required to have its point of closest approach to the
beam axis within $10\un{cm}$ of the interaction point in the beam
direction and within $1\un{cm}$ of the beam axis in the plane
perpendicular to the beam. The polar angle of the track is required
to be within the region $|\cos\theta\,|<0.8$.

The TOF information is used to calculate particle
identification~(PID) probabilities for pion, kaon and proton
hypotheses~\cite{pid_ppbar}. For each track, the particle type yielding the largest
probability is assigned. Here, the momentum of proton is high ($> 1.6\gevc$).
For this high momentum protons and antiprotons, the PID efficiency is about 95\%.
The ratio of kaons to be mis-identified as protons is about 5\%. In this analysis, one charged track is
required to be identified as a proton and the other one as an
antiproton.

The angle between the proton and antiproton ($\theta_{p\bar{p}}$) in
the rest frame of the overall $e^+e^-$ CMS system is required to be greater than
179~degrees. Finally, for both tracks, the absolute difference
between the measured and the expected momentum ({\it e.g.}
$1.637\gevc$ for the $\psi(3770)$ data sample) should be less than
$40\mevc$ (about 3$\sigma$).

\begin{figure}[htbp]
   \centerline{
   \psfig{file=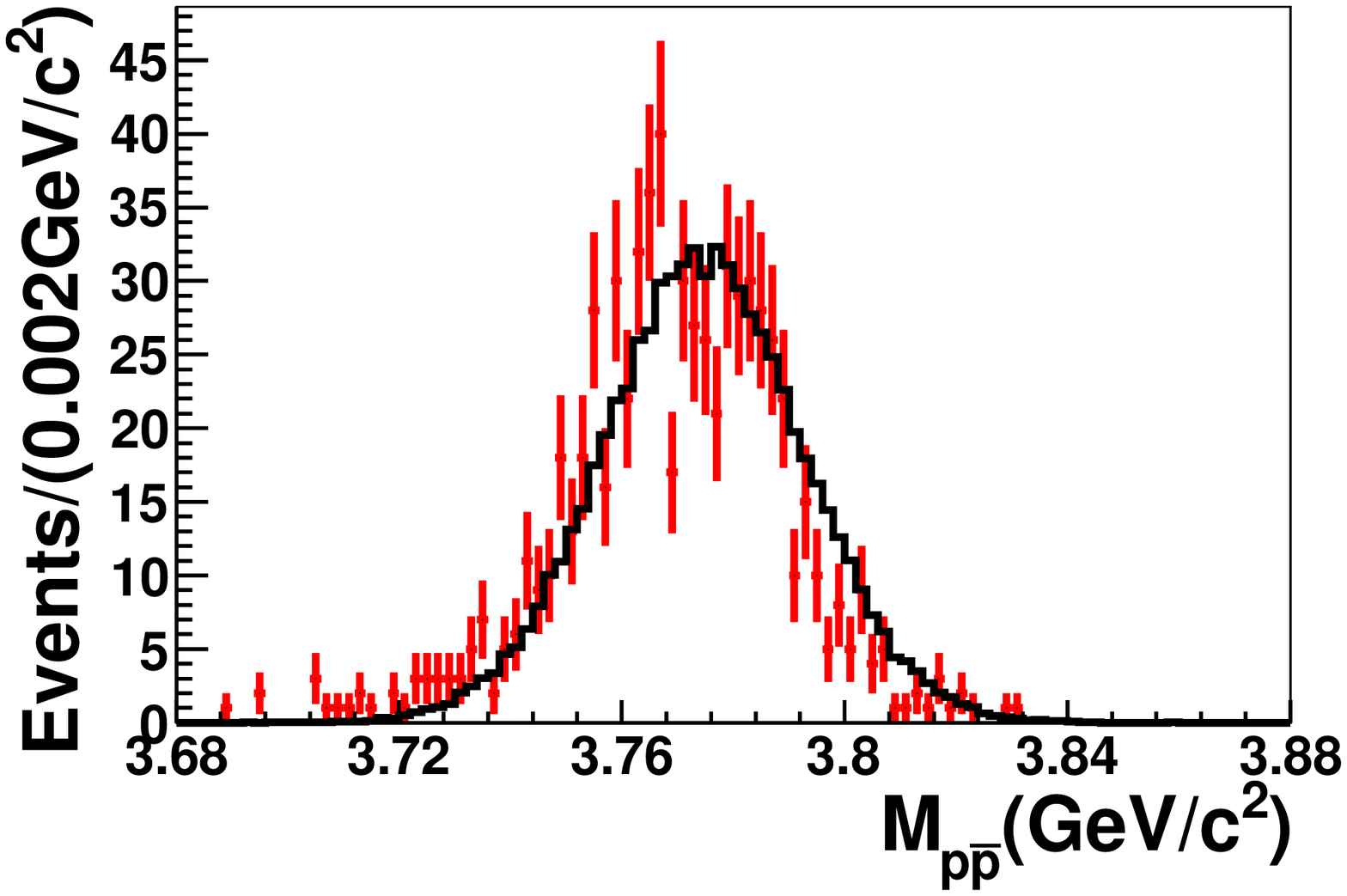, width=4.6cm, angle=0}
              \put(-110,67){(a)}
   \psfig{file=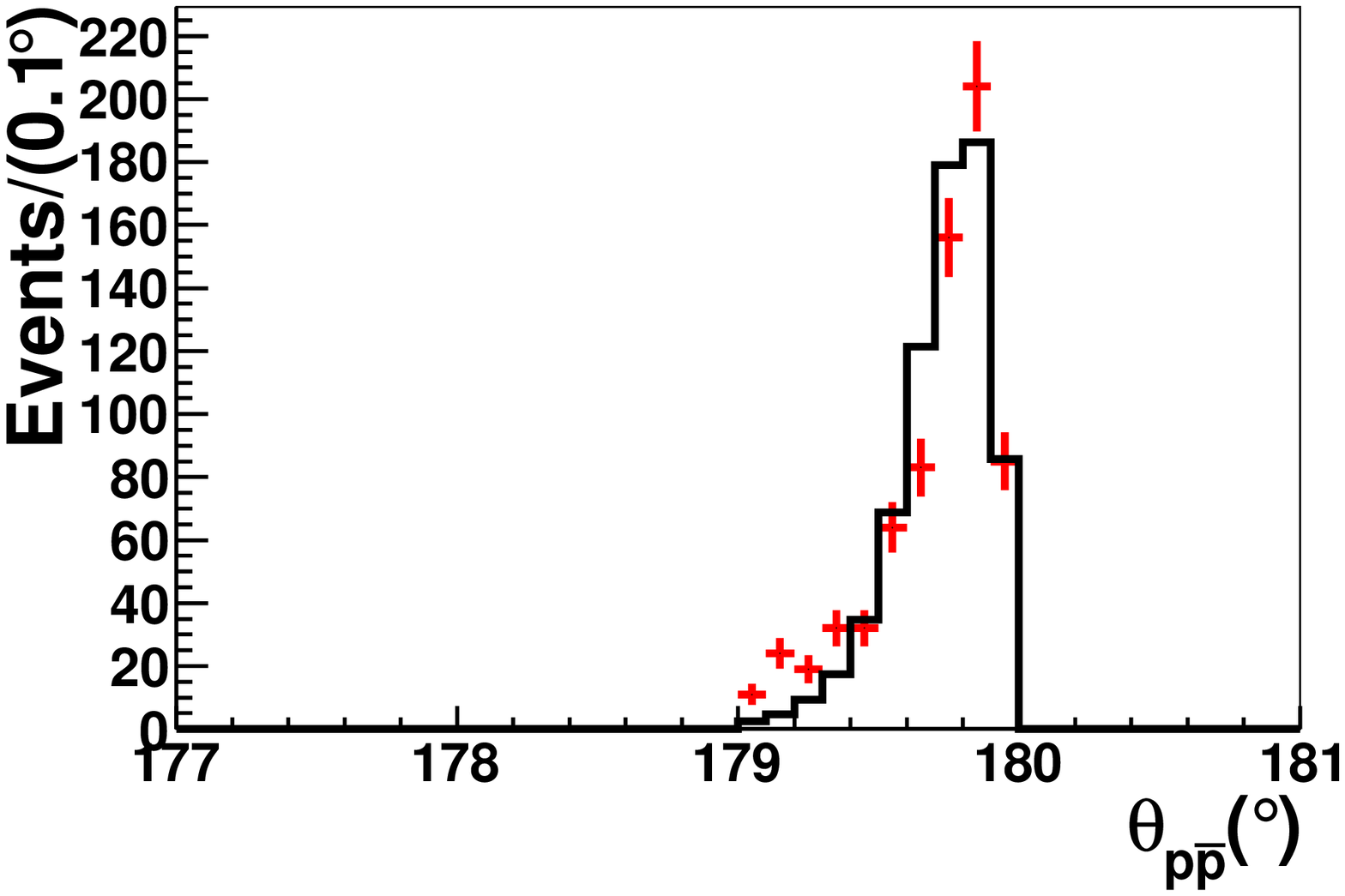, width=4.6cm, angle=0}
              \put(-110,67){(b)}}
   \centerline{
   \psfig{file=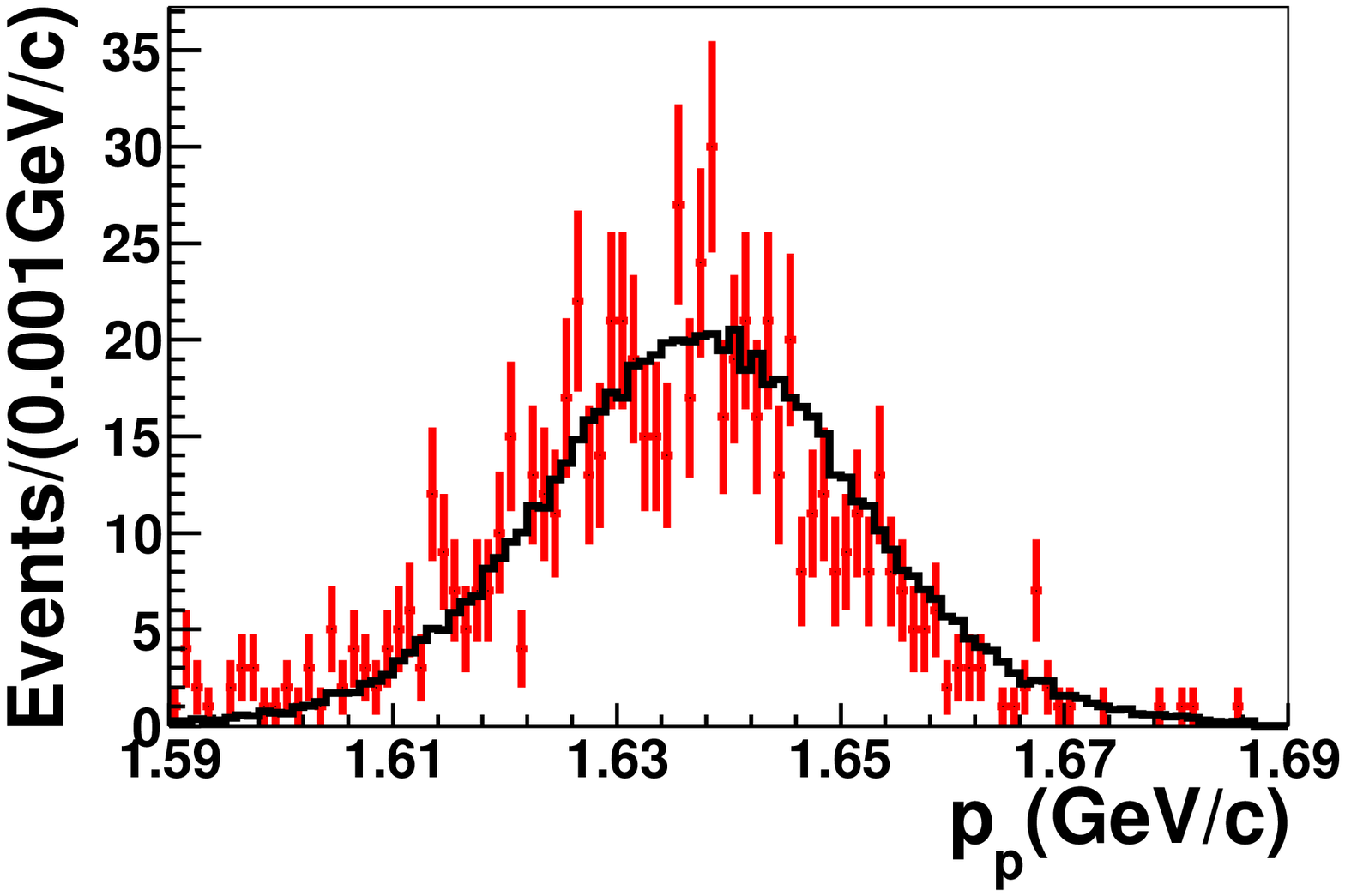, width=4.6cm, angle=0}
              \put(-110,67){(c)}
   \psfig{file=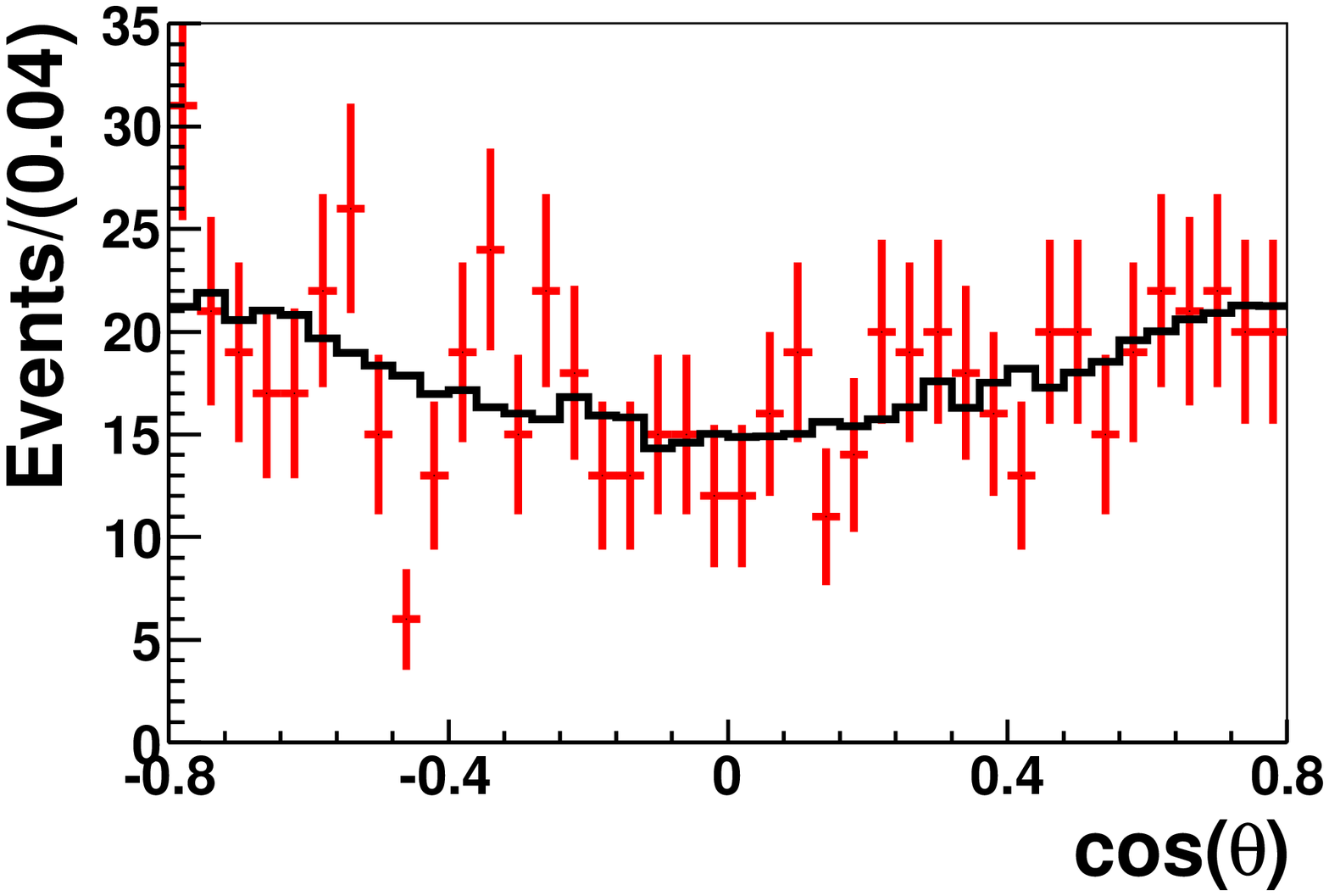, width=4.6cm, angle=0}
              \put(-110,67){(d)}}
 \caption{\label{fig:ppbar_signal}
   Comparisons between experimental and MC simulation data of selected
   $e^+e^-\rightarrow p\bar{p}$ events at $3.773\gev$. (a)~The
   invariant mass of $p\bar{p}$ calculated with raw 4-momenta; (b)~the
   angle between the proton and antiproton~($\theta_{p\bar{p}}$) in the
   rest frame of the overall $e^+e^-$ CMS system; (c)~the magnitude of the proton
   momentum; (d)~the $\cos\theta$ of the proton momentum. The black
   histograms are MC simulations and the red crosses are experimental
   data. }
\end{figure}

After imposing the above event selection criteria, $684\pm26$
candidate events remain from the $\psi(3770)$ data set. Comparisons
between experimental and MC data are plotted in
Fig.~\ref{fig:ppbar_signal}. The MC simulation agrees with the
experimental data. For other data sets, signal events are selected
with similar selection criteria. Signal yields are listed in
Table~\ref{tab:observed_cross_section}.

\begin{table*}
\small
\caption{Summary of results at center-of-mass energies from 3.65 to
$3.90\gev$. $N_{sig}$ is the number of $e^+e^-\rightarrow
p\bar{p}$ events; $\epsilon$ is the detection efficiency; $L$ is the
integrated luminosity; $(1+\delta)_{dressed}$ is the initial state
radiation correction factor without the vacuum polarization
correction; and $\sigma_{obs}$, $\sigma_{dressed}$ and
$\sigma_{Born}$ are the observed cross section, the dressed cross
section and the Born cross section, respectively.
}\label{tab:observed_cross_section}

\begin{tabular}{cccccccc}
\hline\hline
  $\sqrt{s}$(GeV) & $N_{sig}$ &  $\epsilon$(\%) &  $L (\invpb)$ &
  $(1+\delta)_{dressed}$ &$\sigma_{obs}$ (pb) & $\sigma_{dressed} $(pb) & $\sigma_{Born}$(pb)\\
  \hline
 3.650 & $26.0\pm5.1$         &  $62.6\pm0.4$ &  44.5     &0.76 & $0.90\pm0.18\pm0.06$  & $1.19\pm0.24\pm0.08$ &$1.12\pm0.22\pm0.08$\\
 3.748 & $1.0^{+1.8}_{-0.6}$  &  $61.2\pm0.4$ &  3.57      &0.76 & $0.46^{+0.83}_{-0.28}\pm0.03$ & $0.60^{+1.08}_{-0.36}\pm0.04$ & $0.54^{+0.97}_{-0.32}\pm0.04$\\
 3.752 & $3.0^{+2.3}_{-1.9}$  &  $60.8\pm0.4$ &  6.05      &0.76 & $0.82^{+0.63}_{-0.52}\pm0.06$ & $1.07^{+0.82}_{-0.68}\pm0.08$ & $0.96^{+0.74}_{-0.61}\pm0.07$\\
 3.755 & $4.0^{+2.8}_{-1.7}$  &  $61.7\pm0.4$ &  7.01      &0.77 & $0.93^{+0.65}_{-0.39}\pm0.06$ & $1.21^{+0.85}_{-0.51}\pm0.09$ & $1.09^{+0.76}_{-0.46}\pm0.08$\\
 3.760 & $4.0^{+2.8}_{-1.7}$  &  $62.4\pm0.4$ &  8.65      &0.77 & $0.74^{+0.52}_{-0.32}\pm0.05$ & $0.96^{+0.67}_{-0.41}\pm0.07$ & $0.87^{+0.61}_{-0.37}\pm0.06$\\
 3.766 & $0.0^{+1.3}_{-0.0}$    &  $62.4\pm0.4$ &  5.57      &0.79 & $0.00^{+0.37}_{-0.00}$ ($<0.70$) & $0.00^{+0.47}_{-0.00}$ ($<0.89$) & $0.00^{+0.43}_{-0.00}$ ($<0.81$)\\
 3.772 & $0.0^{+1.3}_{-0.0}$    &  $62.5\pm0.4$ &  3.68      &0.80 & $0.00^{+0.56}_{-0.00}$ ($<1.06$) & $0.00^{+0.70}_{-0.00}$ ($<1.33$) & $0.00^{+0.64}_{-0.00}$ ($<1.20$)\\
 3.773 & $684\pm26$       &  $62.3\pm0.4$ &  2917   &0.80 & $0.38\pm0.01\pm0.03$  & $0.47\pm0.02\pm0.04$ &$0.43\pm0.02\pm0.03$\\
 3.778 & $0.0^{+1.3}_{-0.0}$    &  $62.6\pm0.4$ &  3.61      &0.78 & $0.00^{+0.57}_{-0.00}$ ($<1.08$) & $0.00^{+0.74}_{-0.00}$ ($<1.39$) & $0.00^{+0.66}_{-0.00}$ ($<1.25$)\\
 3.784 & $0.0^{+1.3}_{-0.0}$    &  $62.4\pm0.4$ &  4.57      &0.75 & $0.00^{+0.45}_{-0.00}$ ($<0.85$) & $0.00^{+0.60}_{-0.00}$ ($<1.14$) & $0.00^{+0.54}_{-0.00}$ ($<1.02$)\\
 3.791 & $1.0^{+1.8}_{-0.6}$  &  $62.1\pm0.4$ &  6.10      &0.74 & $0.26^{+0.48}_{-0.16}\pm0.02$ & $0.35^{+0.64}_{-0.21}\pm0.02$ & $0.32^{+0.57}_{-0.19}\pm0.02$\\
 3.798 & $3.0^{+2.3}_{-1.9}$  &  $61.9\pm0.4$ &  7.64      &0.75 & $0.63^{+0.49}_{-0.40}\pm0.04$ & $0.85^{+0.65}_{-0.54}\pm0.06$ & $0.77^{+0.59}_{-0.48}\pm0.05$\\
 3.805 & $1.0^{+1.8}_{-0.6}$  &  $61.5\pm0.4$ &  4.34      &0.75 & $0.37^{+0.67}_{-0.22}\pm0.03$ & $0.50^{+0.90}_{-0.30}\pm0.04$ & $0.45^{+0.81}_{-0.27}\pm0.03$\\
 3.810 & $20.0\pm4.5$         &  $62.4\pm0.4$ &  52.60     &0.75 & $0.61\pm0.14\pm0.04$ & $0.81\pm0.18\pm0.06$ & $0.73\pm0.16\pm0.05$\\
 3.819 & $1.0^{+1.8}_{-0.6}$  &  $61.4\pm0.4$ &  1.05      &0.75 & $1.55^{+2.79}_{-0.93}\pm0.11$ & $2.06^{+3.70}_{-1.23}\pm0.14$ & $1.85^{+3.34}_{-1.11}\pm0.13$\\
 3.900 & $12.0^{+4.3}_{-3.2}$ &  $61.7\pm0.4$ &  52.61     &0.76 & $0.37^{+0.13}_{-0.10}\pm0.03$ & $0.49^{+0.17}_{-0.13}\pm0.03$ & $0.44^{+0.16}_{-0.12}\pm0.03$\\
\hline \hline

\end{tabular}
\end{table*}

\section{Background estimation}

Background from ISR to the lower lying $\psi(3686)$ resonance, which
is not taken into account in the ISR correction procedure, is
estimated with a sample of MC-simulated data. The number of expected
background events from this process is 0.1 and is neglected in this
analysis.

Background from $\psi(3770)\rightarrow D\bar{D}$ is estimated with
an inclusive MC sample and can also be neglected. Exclusive
channels, such as $e^+e^- \rightarrow K^+K^-$, $\mu^+\mu^-$,
$\tau^+\tau^-$, $p\bar{p}\pi^0$, $p\bar{p}\gamma$ are also studied. The
total background contribution is estimated to be 0.4 events, which is
equivalent to a contamination ratio of 0.06\%. Contributions from
decay channels with unmeasured branching fractions for the
$\psi(3770)$ are estimated by the branching fractions of the
corresponding decay channels of $\psi(3686)$. These background
contributions from unmeasured decay modes are taken into account in
the systematic uncertainty~(0.06\%) instead of being subtracted
directly.

The data set at $3.65\gev$ contains a contribution from the
$\psi(3686)$ tail, whose cross section is estimated to be
$0.136\pm0.012\nb$~\cite{BES_nonDDbar_4}. The normalized contribution
from this tail, 0.89 events, is also
statistically subtracted from the raw signal yield.

\section{Determination of cross sections}

The observed cross sections at the center-of-mass energies
$\sqrt{s}=3.65$, $3.773\gev$ and the fourteen different energy
points in the vicinity of the $\psi(3770)$ resonance are determined
according to $\sigma = \frac{N_{sig}}{\epsilon L}$, where $\epsilon$
is the detection efficiency determined from MC simulation and $L$ is
the integrated luminosity for each energy point. The observed cross
sections are listed in Table~\ref{tab:observed_cross_section}. For
energy points with no significant signal, upper limits on the cross
section at 90\% C.L. are given using the Feldman-Cousins method from
Ref.~\cite{statistics_paper}.

The observed cross section of $e^+e^-\rightarrow p\bar{p}$ contains
the lowest order Born cross section and some higher order
contributions. The BaBar Collaboration~\cite{babar_G,
babar_G_with_sa} has taken into account bremsstrahlung, $e^+e^-$
self-energy and vertex corrections in their radiative correction.
Vacuum polarization is included in their reported cross section.
This corrected cross section, which is the sum of the Born cross
section and the contribution of vacuum polarization, is called the
dressed cross section. In order to use the BaBar measurements of
$\sigma(e^+e^-\rightarrow p\bar{p})$~\cite{babar_G, babar_G_with_sa}
in our investigation, a radiative correction is performed to
calculate the dressed cross section using the method described in
Refs.~\cite{isr_1, isr_2}. With the observed cross sections as our
initial input, a fit to the line-shape
equation~(Eq.~(\ref{eq:sigma_tot_iso_spin})) is performed
iteratively. At each iteration, the ISR correction factors are
calculated and the dressed cross sections are updated. The
calculation converges after a few iterations~($\sim$ 5). The dressed
cross section at each data point is listed in
Table~\ref{tab:observed_cross_section}. As a reference, the Born
cross sections are also calculated and given in
Table~\ref{tab:observed_cross_section}. The Born cross section
around 3.773 GeV is in excellent agreement with a previous
measurement obtained with CLEO data~\cite{Kamal_cleoc}.

\section{Fit to the cross section}

To extract the $\psi(3770)\rightarrow p\bar{p}$ cross section, the
total cross section as a function of $\sqrt{s}$ is constructed and a
fit to the measured values is performed. As discussed in the
introduction, the measured cross section is composed of three
contributions: the three-gluon resonant process~($A_{3g}$), the
one-photon resonant process~($A_{\gamma}$) and the non-resonant
process~($A_{con}$). For the exclusive light hadron decay of the
$\psi(3770)$, the contribution of the electromagnetic process
$A_{\gamma}$ is negligible compared to that of the three-gluon
strong interaction $A_{3g}$~\cite{P_wang_A3g_Agamma}. The resonant
amplitude can then be written as $A_{\psi}\equiv A_{3g} + A_{\gamma}
\sim A_{3g}$. Finally, the total cross section can be constructed
with only two amplitudes, $A_{\psi}$ and $A_{con}$,
\begin{equation} \label{eq:sigma_tot_iso_spin}
\begin{aligned}
&\sigma(s) = |A_{con} + A_{\psi}e^{i\phi}|^2\\
&=\left|\sqrt{\sigma_{con}(s)}+\sqrt{\sigma_\psi}
\frac{m_\psi\Gamma_\psi}{s-m_\psi^2+im_\psi\Gamma_\psi}e^{i\phi}\right|^2,\\
\end{aligned}
\end{equation}
where $m_\psi$ and $\Gamma_\psi$ are the mass and width of the
$\psi(3770)$~\cite{pdg}, respectively; $\phi$ describes the phase
angle between the continuum and resonant amplitudes, which is a free
parameter to be determined in the fit; and $\sigma_\psi$ is the
resonant cross section, which is also a free parameter.

The continuum cross section, $\sigma_{con}$, has been measured by
many experiments~\cite{babar_G, babar_G_with_sa, bes2_G,
babar_G_old}. In Ref.~\cite{bes2_G} from the BESII Collaboration,
$\sigma_{con}$ was measured from $2$ to $3.07\gev$, and is
well-described with an $s$ dependence according to
\begin{equation} \label{eq:sigma_con}
\sigma_{con}(s)=\frac{4\pi\alpha^2v}{3s}\left(1+\frac{2m_p^2}{s}\right)|G(s)|^2,
\end{equation}
\begin{equation} \label{eq:form_factor}
|G(s)|=\frac{C}{s^2\ln^2(s/\Lambda^2)}.
\end{equation}

Here $\alpha$ is the fine-structure constant; $m_p$ is the nominal
proton mass; $v$ is the proton velocity in the $e^+e^-$ rest frame;
$G(s)$ is the effective proton form
factor~\cite{babar_G_old}; $\Lambda = 0.3\gev$ is the QCD scale
parameter; and $C$ is a free parameter.



\begin{figure}[htbp]
   \centerline{
   \psfig{file=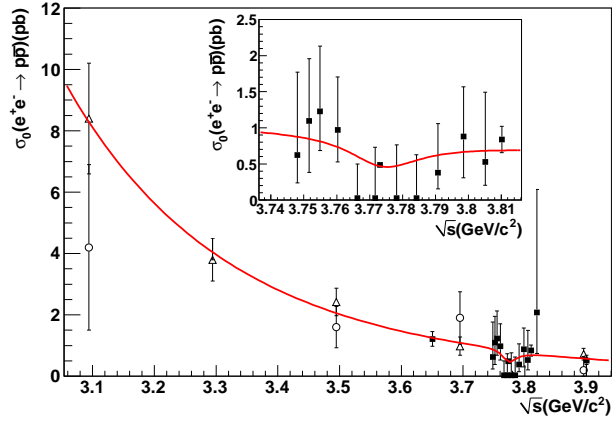, width=9cm, angle=0}}

            \caption{\label{fig:fit} Fit to the dressed cross section of $e^+e^-\rightarrow p\bar{p}$ as
a function of center-of-mass energy. The red dashed line shows the
fit curve. The solid square points with error bars are from BESIII.
The open circles are from the BaBar measurements of
Ref.~\cite{babar_G}, and the open triangles from
Ref.~\cite{babar_G_with_sa}. The inset shows a zoom of the region in
the vicinity of the $\psi(3770)$.}

\end{figure}

The dressed cross sections in
Table~\ref{tab:observed_cross_section}, together with the BaBar
measurements of the cross sections between $3$ and $4\gev$, are fitted
with Eq.~(\ref{eq:sigma_tot_iso_spin}). In this fit, 26~data points
are considered: 16 points from this investigation by BESIII, 5
points from Ref.~\cite{babar_G} and 5 points from
Ref.~\cite{babar_G_with_sa}. The free parameters are the phase angle
$\phi$, the resonant cross section $\sigma_\psi$, and $C$ from the
form factor describing the contribution of the continuum.
Fig.~\ref{fig:fit} shows the data points and the fit result.

The fit yields a $\chi^2/ndf$ of $13.4/23$. Two solutions are found
with the same $\chi^2$ and the same parameter $C$ of $(62.0\pm2.3)
\un{GeV^4}$. Two solutions are found because the cross section in
Eq.~(\ref{eq:sigma_tot_iso_spin}) is constructed with the square of
two amplitudes. This multi-solution problem has been explained in
Ref.~\cite{zhuk_ijmp_multi_solution}. A dip indicating destructive
interference is seen clearly in the fit (the red solid line in
Fig.~\ref{fig:fit}). The first solution for the cross section is
$\sigma_{dressed}(e^+e^-\rightarrow\psi(3770)\rightarrow p\bar{p}) =
(0.059 ^{+0.070}_{-0.020})\pb$ with a phase angle $\phi =
(255.8^{+39.0}_{-26.6})^\circ$ ($<0.166\pb$ at the 90\% C.L.). The second
solution is $\sigma_{dressed}(e^+e^-\rightarrow\psi(3770)\rightarrow
p\bar{p}) = (2.57 ^{+0.12}_{-0.13})\pb$ with a phase angle $\phi =
(266.9^{+6.1}_{-6.3})^\circ$.

For comparison, an alternative fit with only the BESIII data points
is performed. Two solutions are found with the same $\chi^2/ndf$ of
$6.8/13$ and the same parameter $C$ of $(62.6\pm4.1) \un{GeV}^4$. The
first solution for the cross section is
$\sigma_{dressed}(e^+e^-\rightarrow\psi(3770)\rightarrow p\bar{p}) =
(0.067 ^{+0.088}_{-0.034})\pb$ with a phase angle $\phi =
(253.8^{+40.7}_{-25.4})^\circ$. The second solution is
$\sigma_{dressed}(e^+e^-\rightarrow\psi(3770)\rightarrow p\bar{p}) =
(2.59 \pm 0.20)\pb$ with a phase angle $\phi =
(266.4\pm6.3)^\circ$. These two solutions agree with those from the
previous fit, but have larger uncertainties.

Table~\ref{tab:cross_section_3770_ppbar} shows a summary of the fit
results, where the first error is from the fit and the second error
is from the correlated systematic uncertainties.

\begin{table}
\centering \caption{ Summary of the extracted results for different
solutions of the fit. Upper limits are determined at 90\%
C.L.}\label{tab:cross_section_3770_ppbar}
\begin{tabular}{ccc}
\hline\hline  Solution &  $\sigma^{dressed}_{(\psi(3770)\rightarrow
p\bar{p})}$ (pb) & $\phi$ ($^\circ$)
\\\hline \multirow{2}{*}{(1)}  & $0.059^{+0.070}_{-0.020}\pm0.012$  &
\multirow{2}{*}{$255.8^{+39.0}_{-26.6}\pm4.8$}
\\ & ($<$ 0.166 at 90\% C.L.) &
\\\hline (2) & $2.57^{+0.12}_{-0.13}\pm0.12$ & $266.9^{+6.1}_{-6.3}
\pm0.9$
\\\hline \hline
\end{tabular}

\end{table}

\section{Systematic uncertainty study}

The sources of systematic uncertainty in the cross section
measurements are divided into two categories: uncorrelated and
correlated uncertainties between different energy points. The former
includes only the statistical uncertainty in the MC simulated
samples (0.4\%), which can be directly considered in the fit.  The
latter refers to the uncertainties that are correlated among
different energy points, such as the tracking (4\% for two charged
tracks), particle identification (4\% for both proton and
antiproton), and integrated luminosity. The integrated luminosity
for the data was measured by analyzing large angle Bhabha
scattering events~\cite{Lumi} and has a total uncertainty of 1.1\%
at each energy point.

To estimate the uncertainty from the radiative corrections, a
different correction procedure using the structure-function
method~\cite{isr_structure_function} is applied, and the difference
in results from these two correction procedures~(2\%) is taken as
the uncertainty. To investigate the impact of the possible
inconsistency of the MC simulation and experimental data, an
alternative MC simulated sample is generated with a different proton
momentum resolution (15\% better than the previous MC sample), and
the change in the final results~(1.4\%) is taken as the uncertainty.

In addition, the uncertainty on the
reconstruction efficiency from the unmeasured angular
distribution of the proton in the rest frame of the overall $e^+e^-$ CMS system is
also studied. According to hadron helicity conservation, the angular
distribution of $\psi\rightarrow p\bar{p}$ can be expressed as
$\frac{dN}{dcos\theta} \propto 1+\alpha \cos^2\theta$, where
$\theta$ is the angle between the proton and the positron beam
direction in the center-of-mass system. The theoretical value of
$\alpha=0.813$~\cite{carimalo_alpha} is used to produce the MC
simulated sample in this analysis. In the case of
$\psi(3686)\rightarrow p\bar{p}$, the mean value of $\alpha$
measured by E835 (0.67$\pm$0.16)~\cite{alpha_psi2s_e385} differs by
$0.13$ from the theoretical value of $0.80$. To obtain a conservative
uncertainty, an alternative MC simulated sample with $\alpha=0.683$
is used and the difference in the results~(1.0\%) is taken as the
uncertainty. The uncertainty from the angle cut between the proton
and antiproton is investigated by varying the angle cut (from 178.9
to 179.5 degrees) and the difference~(2.2\%) is taken as the
uncertainty.

All of the above sources of uncertainty are applied to the observed
cross section at each energy point. The total systematic uncertainty
of the individual energy points is 6.7\%.

The systematic uncertainties on the parameters extracted from the
fit, such as $\sigma^{dressed}_{(\psi(3770)\rightarrow p\bar{p})}$
and the phase angle $\phi$, are estimated by the ``offset
method"~\cite{correlated_error}, in which the error propagation is
determined from shifting the data by the aforementioned correlated
uncertainties and adding the deviations in quadrature. In addition,
a $1\mev$ uncertainty for the beam energy measurements of all the data
points is considered in the fit.

\section{Summary and Discussion}

Using $2917\invpb$ of data collected at $3.773\gev$,
$44.5\invpb$ of data collected at $3.65\gev$ and data
collected during a $\psi(3770)$ line-shape scan with the BESIII
detector, the reaction $e^+e^-\rightarrow p\bar{p}$ has been
studied. To extract the cross section of $e^+e^-\rightarrow
\psi(3770) \rightarrow p\bar{p}$, a fit, taking into account the
interference of resonant and continuum amplitudes, is performed. In
this investigation, the measured cross sections of
$e^+e^-\rightarrow p\bar{p}$ from the BaBar experiment are included
in a simultaneous fit to put more constraints on the continuum
amplitude. The dressed cross section of
$e^+e^-\rightarrow\psi(3770)\rightarrow p\bar{p}$ is extracted from
the fit and shown in Table~\ref{tab:cross_section_3770_ppbar}.

With the obtained dressed cross section of $e^+e^-\rightarrow
\psi(3770)\rightarrow p\bar{p}$, the branching fraction
$B_{\psi(3770)\rightarrow p\bar{p}}$ is determined to be
$(7.1^{+8.6}_{-2.9})\times10^{-6}$ or $(3.1\pm0.3)\times10^{-4}$, by dividing the
dressed cross section of $e^+e^-\rightarrow
\psi(3770)$~\cite{CLEO_nonDDbar}. Even the
larger solution has a relatively small branching fraction comparing
to the large total non-$D\bar{D}$ branching fraction. Thus, the
$p\bar{p}$ channel alone cannot explain the large non-$D\bar{D}$ branching
fraction from BESII.

Using the branching fraction of $\psi(3770)\rightarrow p\bar{p}$,
the cross section of its time reversed reaction $p\bar{p}\rightarrow
\psi(3770)$ can be estimated using the Breit-Wigner
formula~\cite{pdg}:

\begin{equation} \label{eq:cross_section_ppbar_to_3770}
\sigma_{p\bar{p}\rightarrow\psi(3770)}(s)=
\frac{4\pi(2J+1)}{(s-4m_p^2)} \frac{B_{\psi(3770)\rightarrow
p\bar{p}}}{1+[2(\sqrt{s}-M_{\psi})/\Gamma_{\psi}]^2}
\end{equation}

where $M_\psi$ and $\Gamma_\psi$ are the mass and width of the
$\psi(3770)$ resonance, J is the spin of the $\psi(3770)$, and $m_p$
is the proton mass. For the condition $\sqrt{s}=M_\psi$, the cross
section $\sigma(p\bar{p}\rightarrow \psi(3770))$ is estimated to be
either $(9.8^{+11.8}_{-3.9})\nb$ ($<27.5\nb$ at 90\% C.L.) or
$(425.6^{+42.9}_{-43.7})\nb$.

The future $\rm \bar{P}ANDA$ (anti-Proton ANnihilations at
DArmstadt) experiment is one of the key projects at the Facility for
Antiproton and Ion Research (FAIR), which is currently under
construction at GSI, Darmstadt. It will perform precise studies of
antiproton-proton annihilations with various internal proton or
nuclear targets and an anti-proton beam in the momentum range from
1.5 GeV/c to 15 GeV/c. In $\rm \bar{P}ANDA$, a detailed
investigation of the charmonium spectrum and the open charm channels
is foreseen. For this physics program, it is important to obtain
experimental information on the so far unknown open charm cross
sections, both to evaluate luminosity requirements and to design
detector. Theoretical estimations vary with several orders of
magnitude~\cite{ppbar_ccbar_theory_1, ppbar_ccbar_theory_2,
ppbar_ccbar_theory_3, ppbar_ccbar_theory_4, ppbar_ccbar_theory_5,
ppbar_ccbar_theory_6, ppbar_ccbar_theory_7, ppbar_ccbar_theory_8,
ppbar_ccbar_theory_9}. In the physics performance report for $\rm
\bar{P}ANDA$~\cite{panda_physics_report}, the $D\bar{D}$ production
cross section is estimated to be $6.35\nb$, with the unknown branching
ratio of $\psi(3770)\rightarrow p\bar{p}$ scaled
from the known
ratio of $J/\psi \rightarrow p\bar{p}$. In this paper, the cross
section of $\sigma(p\bar{p}\rightarrow \psi(3770))$
has been determined.
As the first charmonium state above the $D\bar{D}$ threshold,
$\psi(3770)$ could be used as a source of open charm production.

In this paper, two solutions on the cross section of
$\sigma(p\bar{p}\rightarrow \psi(3770))$ are obtained. It is
impossible to distinguish these two solutions with our data. The
first solution, $(9.8^{+11.8}_{-3.9}) \nb$, is compatible with a
simple scaling from $J/\psi$ used in the $\rm \bar{P}ANDA$ physics
performance report. The second solution, with the cross section of
$(425.6^{+42.9}_{-43.7}) \nb$, is two order of magnitudes larger.


\section{Acknowledgement}
The BESIII collaboration thanks the staff of BEPCII and the
computing center for their strong support. This work is supported in
part by the Ministry of Science and Technology of China under
Contract No. 2009CB825200; Joint Funds of the National Natural
Science Foundation of China under Contracts Nos. 11079008, 11179007,
U1332201; National Natural Science Foundation of China (NSFC) under
Contracts Nos. 10625524, 10821063, 10825524, 10835001, 10935007,
11125525, 11235011; the Chinese Academy of Sciences (CAS)
Large-Scale Scientific Facility Program; CAS under Contracts Nos.
KJCX2-YW-N29, KJCX2-YW-N45; 100 Talents Program of CAS; German
Research Foundation DFG under Contract No. Collaborative Research
Center CRC-1044; Istituto Nazionale di Fisica Nucleare, Italy;
Ministry of Development of Turkey under Contract No.
DPT2006K-120470; U. S. Department of Energy under Contracts Nos.
DE-FG02-04ER41291, DE-FG02-05ER41374, DE-FG02-94ER40823,
DESC0010118; U.S. National Science Foundation; University of
Groningen (RuG) and the Helmholtzzentrum fuer Schwerionenforschung
GmbH (GSI), Darmstadt; WCU Program of National Research Foundation
of Korea under Contract No. R32-2008-000-10155-0.


\begin{thebibliography}{99}


\bibitem{MARK}
  P. A. Rapidis {\em et al.},
  Phys. Rev. Lett. {\bf 39} (1978) 526.

\bibitem{DELCO}
  W. Bacino {\em et al.},
  Phys. Rev. Lett. {\bf 40} (1978) 671 .

\bibitem{BES_nonDDbar_1}
  M. Ablikim {\em et al.} [BES Collaboration],
  Phys. Lett. B {\bf 641} (2006) 145.

\bibitem{BES_nonDDbar_2}
  M. Ablikim {\em et al.} [BES Collaboration],
  Phys. Rev. Lett. {\bf 97} (2006) 121801.

\bibitem{BES_nonDDbar_3}
  M. Ablikim {\em et al.} [BES Collaboration],
  Phys. Lett. B {\bf 659} (2007) 74.

\bibitem{BES_nonDDbar_4}
  M. Ablikim {\em et al.} [BES Collaboration],
  Phys. Rev. D {\bf 76} (2007) 122002.

\bibitem{CLEO_nonDDbar}
  D. Besson {\em et al.} [CLEO Collaboration],
  Phys. Rev. Lett. {\bf 96} (2006) 092002.

\bibitem{interfere_wangp_1}
  P. Wang, C. Yuan, X. Mo, and D. Zhang,
  hep-ph/0212139 (2002).

\bibitem{non_DDbar_exclucive_1}
  M. Ablikim {\em et al.} [BESIII Collaboration],
  Phys. Rev. D {\bf 87} (2013) 112011.

\bibitem{non_DDbar_exclucive_2}
  G. S. Adams {\em et al.} [CLEO Collaboration],
  Phys. Rev. D {\bf 73} (2006) 012002.

\bibitem{ppbarpi0_matthias}
  M. Ablikim {\em et al.} [BESIII Collaboration], arXiv:1406.2486 [hep-ex] (2014).

\bibitem{BESIII_BEPCII}
  M. Ablikim {\em et al.} [BESIII Collaboration],
  Nucl. Instrum. Meth. A \textbf{614} (2010) 345.

\bibitem{Lumi}
  M. Ablikim {\em et al.} [BESIII Collaboration],
  Chinese Physics C {\bf 37} (2013) 123001.

\bibitem{geant4}
  S.~Agostinelli {\it et al.}  [GEANT4 Collaboration],
  Nucl.\ Instrum.\ Meth.\ A {\bf 506} (2003) 250.


\bibitem{generator}
  R. G. Ping,
  Chinese Physics C {\bf 32} (2008) 599.

\bibitem{isr_1}
  H. Hu, X. Qi, G. Huang, and Z. Zhao, High Energy Physics and
  Nuclear Physics {\bf 25} (2001) 701.

\bibitem{isr_2}
  C. Edwards {\it et al.},
  SLAC-PUB-5160 (1990).

\bibitem{phokhara}
  H. Czyz {\it et al.},
  Eur. Phys. J. C {\bf 35} (2004) 527.

\bibitem{pid_ppbar}
  M. Ablikim {\em et al.} [BESIII Collaboration],
   Phys. Rev. D {\bf 86} (2012) 032014.

\bibitem{statistics_paper}
  G. J. Feldman and R. D. Cousins,
  Phys. Rev. D {\bf 57} (1998) 3873.

\bibitem{babar_G}
  J. P. Lees {\em et al.} [BABAR Collaboration],
  Phys. Rev. D {\bf 87} (2013) 092005.

\bibitem{babar_G_with_sa}
  J. P. Lees {\em et al.} [BABAR Collaboration],
  Phys. Rev. D {\bf 88} (2013) 072009.

\bibitem{Kamal_cleoc}
  Kamal K. Seth {\em et al.}, Phys. Rev. Lett. {\bf 110} (2013)
022002.

\bibitem{P_wang_A3g_Agamma}
  P. Wang, X. H. Mo, and C. Z. Yuan,
  International Journal of Modern Physics A {\bf 21} (2006) 5163.

\bibitem{pdg}
  J. Beringer {\it et al.} [Particle Data Group],
  Phys. Rev D {\bf 86} (2012) 010001.

\bibitem{bes2_G}
  M. Ablikim {\em et al.} [BES Collaboration],
  Phys. Lett. B {\bf 630} (2005) 14.

\bibitem{babar_G_old}
  B. Aubert {\em et al.} [BABAR Collaboration],
  Phys. Rev. D {\bf 73} (2006) 012005.

\bibitem{zhuk_ijmp_multi_solution}
  K. Zhu {\em et al.},
  Int. J.Mod. Phys. A {\bf 26} (2011) 4511.




\bibitem{isr_structure_function}
  E. A. Kuraev and V. S. Fadin, Sov. J. Nucl. Phys {\bf 41} (1985) 779.

\bibitem{carimalo_alpha}
  C. Carimalo, Int. J. Mod. Phys. A {\bf 2} (1987) 249.

\bibitem{alpha_psi2s_e385}
  M.Ambrogiani {\em et al.} [Fermilab E835 Collaboration],
  Phys. Lett. B {\bf 610} (2005) 177.

\bibitem{correlated_error}
  M. Botje,
  Journal of Physics G: Nuclear and Particle Physics {\bf 28} (2002) 779.

\bibitem{ppbar_ccbar_theory_1}
  A. Khodjamirian, C. Klein, Th. Mannel, and Y.M. Wang,
  Eur. Phys. J. A {\bf 48 } (2012) 31.

\bibitem{ppbar_ccbar_theory_2}
  A. I. Titov and B. Kampfer,
  Phys. Rev. C {\bf 78} (2008) 025201.

\bibitem{ppbar_ccbar_theory_3}
  A. I. Titov and B. Kampfer,
  arXiv:1105.3847 [hep-ph].

\bibitem{ppbar_ccbar_theory_4}
  P. Kroll, B. Quadder and W. Schweiger,
  Nucl. Phys. B {\bf 316} (1989) 373.

\bibitem{ppbar_ccbar_theory_5}
  A. T. Goritschnig, P. Kroll and W. Schweiger,
  Eur. Phys. J. A {\bf 42} (2009) 43.

\bibitem{ppbar_ccbar_theory_6}
  E. Braaten and P. Artoisenet,
  Phys. Rev. D {\bf 79} (2009) 114005.

\bibitem{ppbar_ccbar_theory_7}
  J. Haidenbauer and G. Krein,
  Phys. Lett. B {\bf 687} (2010) 314.

\bibitem{ppbar_ccbar_theory_8}
  J. Haidenbauer and G. Krein,
  Few Body Syst. {\bf 50} (2011) 183.

\bibitem{ppbar_ccbar_theory_9}
  B. Kerbikov and D. Kharzeev,
  Phys. Rev. D {\bf 51} (1995) 6103.

\bibitem{panda_physics_report}
    M.~F.~M.~Lutz {\em et al.} [PANDA Collaboration],
    arXiv:0903.3905v1 [hep-ex] (2009).

\end{thebibliography}

\end{document}